\documentclass[pra, showpacs, twocolumn, floatfix]{revtex4}
\usepackage{graphicx}
\usepackage{amsmath, amsfonts, amssymb, bm}
\begin{document}
%%%%%%%%%%%%%%%%%%%%%%%%%%%%%%%%%%%%%%%%%%%%%%%%%%%%%%%%%%%%%%%%%
\title{Optical force acting on strongly driven atoms in free space 
or modified reservoirs}
\author{Mihai A. \surname{Macovei}$^{\dagger}$}
\email{mihai.macovei@mpi-hd.mpg.de}
\affiliation{Max-Planck-Institut f\"{u}r Kernphysik, Saupfercheckweg
1, D-69117 Heidelberg, Germany}
\date{\today}
%%%%%%%%%%%%%%%%%%%%%%%%%%%%%%%%%%%%%%%%%%%%%%%%%%%%%%%%%%%%%%%%%
\begin{abstract}
We investigate the quantum dynamics of a moderately driven two-level 
particle in free space or modified electromagnetic field reservoir. 
Particularly, we calculate the optical force acting on the radiator 
in such an environment. We found that the modified environmental 
reservoir influences significantly the optical force. Very intense 
driving in free space also modifies the maximal force.
\end{abstract}
%%%%%%%%%%%%%%%%%%%%%%%%%%%%%%%%%%%%%%%%%%%%%%%%%%%%%%%%%%%%%%%%%%%
\maketitle
%%%%%%%%%%%%%%%%%%%%%%%%%%%%%%%%%%%%%%%%%%%%%%%%%%%%%%%%%%%%%%%%%%%

%%%%%%%%%%%%%%%%%%%%%%%%%%%%%%%%%%%%%%%%%%%%%%%%%%%%%%%%%%%%%%%%%%%
\section{Introduction}
%%%%%%%%%%%%%%%%%%%%%%%%%%%%%%%%%%%%%%%%%%%%%%%%%%%%%%%%%%%%%%%%%%%
The force acting on a two-level atom in a resonance light field can be estimated 
as follows: in the field of a strong running wave, the atom absorbs a
photon from the light beam and acquires the momentum $\hbar k$ of the photon. 
Respectively, a force $\hbar k \gamma$ acts on the atom, where $2\gamma$ is the 
spontaneous-decay rate of the upper level \cite{kaz,ga,ph}. This force can be even 
stronger in a field of a standing wave. If the atom is accelerated in such a field 
only to a distance of half the wavelength, it acquires an energy greater than in the 
thermal case. The acceleration effect can be substantially enhanced if the frequency 
of one of the opposing waves varies with time. Acceleration of neutral particles was 
achieved in Ref.~\cite{exp_ac}. Furthermore, the scattering rate from a coherent 
stimulated process can be made arbitrarily large by detuning the optical field far from 
resonance and increasing the intensity. When detuned from resonance very large 
accelerations in the $10^{11}$g range have been realized and this process has been 
termed optical Stark deceleration or acceleration \cite{barker}. Thus, the mechanical 
effect of a resonant or non-resonant field on atoms can be considerable. In an another 
context, laser acceleration of charged particles up to GeV energies was obtained as 
well \cite{Leem,YS_CHK}. Previous force studies include dielectrics \cite{diel1,diel2} 
and plasmas \cite{pl1,pl2}.

Another important related issue is the cooling and trapping in laser fields \cite{HS}. 
The experimental feasibility of Bose-Einstein condensation is already history \cite{ket}.
Laser-cooled atoms are used, for example, as frequency standards \cite{fs}, in quantum 
information processing \cite{qp} or in atomic clocks \cite{cl}. Therefore, it is not 
suprising that the optical force received a lot of attention. In particular, cooling of 
atoms with stimulated emission was observed in \cite{cool_exp}. Laser cooling of atoms in 
squeezed vacumm was investigated in \cite{cool_sqv1,cool_sqv2}, respectively. An overview 
on cold atoms and quantum control was given in Ref.~\cite{chu1} while laser cooling of atoms,
ions or molecules by coherent scattering was studied in \cite{chu2}. Adiabatic cooling of 
atoms by an intense standing wave was experimentally achieved in \cite{ad_cool}. Further, 
in Ref.~\cite{tr_mod} trapping and cooling of atoms in a vacuum perturbed in a 
frequency-dependent manner was investigated. Light-pressure cooling of crystals was annalysed
as well \cite{cr}. Stopping atoms with laser light was achieved in \cite{st} while 
collective-emission-induced cooling of atoms in an optical cavity was observed in Ref.~\cite{cav}.
In the radiation field of an optical waveguide, the Rayleigh scattering of photons was 
shown to result in a strongly velocity-dependent force on atoms \cite{dom}. Finally, these 
techniques were exported to other systems such as mesoscopic systems. For instance, the resolved 
sideband laser cooling was used to cool a mesoscopic mechanical resonator to near its 
quantum ground state \cite{mes}.
%%%%%%%%%%%%%%%%%%%%%%%%%%%%%%%%%%%%%%%%%%%%%%%%%%%%%%%%%%%%%%%%%%%%%%%%%%
\begin{figure}[b]
\includegraphics[width=6cm]{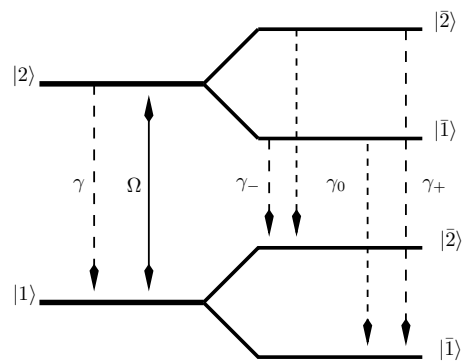}
\caption{\label{fig-1} The scematic picture shown the involved energy levels of 
a two-level atom. Here $\Omega$ is the Rabi frequency while $\gamma$ is the bare 
state spontaneous decay rate. The dressed-state decay rates $\{\gamma_{0},\gamma_{\pm}\}$
are different for a very intense laser field in free space or for a modified environmental 
electromagnetic field reservoir such as cavities.}
\end{figure}
%%%%%%%%%%%%%%%%%%%%%%%%%%%%%%%%%%%%%%%%%%%%%%%%%%%%%%%%%%%%%%%%%%%%%%%%%%%

In this article we investigate the optical force acting on two-level atoms in moderately intense 
running laser fields in free space or modified electromagnetic field (EMF) reservoirs like 
low quality optical cavities. We show that modified reservoirs lead to a 
significant enhancement of optical forces. In contrast, very intense driving in free space 
contributes to a maximal force which is slightly smaller than for moderate pumping. This is 
somehow surprising as one may expect the force to be larger for bigger intensities. Note also 
that modified environmental reservoirs are responsible for recovering of the interference 
pattern \cite{melk}, enhanced squeezing \cite{lsq}, population inversion \cite{john} or 
thresholdless lasing \cite{gxl}.

The article is organized as follows. In Section II we descibe the system of interest and obtain 
an expression for the optical force acting on strongly driven two-level atoms in free space 
or modified reservoirs. Section III deals with discussions of the obtained results. The 
Summary is given in Section IV.

%%%%%%%%%%%%%%%%%%%%%%%%%%%%%%%%%%%%%%%%%%%%%%%%%%%%%%%%%%%%%%%%%%%
\section{Quantum dynamics in moderately strong laser fields and modified EMF reservoir}
%%%%%%%%%%%%%%%%%%%%%%%%%%%%%%%%%%%%%%%%%%%%%%%%%%%%%%%%%%%%%%%%%%%
We proceed by briefly introducing the main steps of the analytical formalism involved 
and then rigorously describing the obtained results. The Hamiltonian characterizing 
the interaction of a two-level particle possessing the frequency $\omega_{0}$ with a 
coherent source of frequency $\omega_{L}$, in a frame rotating at $\omega_{L}$, is \cite{rev}:
$H=H_{0}+H_{L}+H_{F}$, where
%%%%%%%%%%%%%%%%%%%%%%%%%%%%%%%%%%%%%%%%%%%%%%%%%%%%%%%%%%%%%%%%%%%%%%%%%%
\begin{subequations}
\label{Hm}
\begin{align}
& H_{0} = \sum_{k}\hbar(\omega_{k}-\omega_{L})a^{\dagger}_{k}a_{k}+\hbar(\omega_{0}-\omega_{L})S_{z},
\\
& H_{L} = \hbar \Omega(S^{+}e^{i\vec k_{L}\vec r} + S^{-}e^{-i\vec k_{L}\vec r}), 
\\
& H_{F} = i\sum_{k}(\vec g_{k}\cdot \vec d)(a^{\dagger}_{k}S^{-}e^{-i\vec k\vec r}
-a_{k}S^{+}e^{i\vec k\vec r}).
\end{align}
\end{subequations}
%%%%%%%%%%%%%%%%%%%%%%%%%%%%%%%%%%%%%%%%%%%%%%%%%%%%%%%%%%%%%%%%%%%%%%%%%%
Here $H_{0}$ describes the free Hamiltonians of the electromagnetic field and 
the atomic subsystems, respectively. The interaction between the laser field 
with Rabi frequency $\Omega$ and wave vector $\vec k_{L}$, and the two-level 
radiator is given by $H_{L}$. $H_{F}$ characterizes the interaction of the atom 
with the surrounded electromagnetic field reservoir and in free space
$\vec g_{k}=\sqrt{2\pi\hbar\omega_{k}/V}\vec e_{\lambda}$ with $\vec e_{\lambda}$ 
being the photon polarization vector with $\lambda=1,2$ and $V$ is the EMF 
quantization volume. Further, $S^{+}=|2\rangle \langle 1|~[S^{-}=|1\rangle \langle 2|]$ 
describes the excitation [deexcitation] of the two-level particle at position $\vec r$ 
and obeys the commutation relations for su(2) algebra: $[S^{+},S^{-}]=2S_{z}$ and 
$[S_{z},S^{\pm}]=\pm S^{\pm}$. Here $S_{z}=(|2\rangle \langle 2|- |1\rangle \langle 1|)/2$
is the bare-state inversion operator. $a^{\dagger}_{k}$ and $a_{k}$ are the creation and 
the annihilation operator of the EMF, respectively, and satisfy the standard bosonic 
commutation relations, i.e., $[a_{k},a^{\dagger}_{k'}]=\delta_{kk'}$, and $[a_{k},a_{k'}]=[a^{\dagger}_{k},a^{\dagger}_{k'}]=0$. 

The optical force acting on two-level particles in a traveling laser wave is 
given by the following expression:
\begin{eqnarray}
\vec F = - \nabla H_{L}(\vec r) = -i\vec k_{L}\hbar\Omega (S^{+}-S^{-}), \label{f}
\end{eqnarray}
where we considered that the particle is located at the origin, i.e. 
$e^{\pm i\vec k_{L}\vec r} \to 1$.

In what follows we shall describe our system using the dressed-states formalism
(see Fig.~\ref{fig-1}):

%%%%%%%%%%%%%%%%%%%%%%%%%%%%%%%%%%%%%%%%%%%%%%%%%%%%%%%%%%%%%%%%%%%%%%%%%%
\begin{subequations}
\label{ds}
\begin{align}
&|1\rangle=\cos\theta |\bar 1\rangle + \sin\theta|\bar 2\rangle, \\
&|2\rangle=\cos\theta |\bar 2\rangle - \sin\theta|\bar 1\rangle,
\end{align}
\end{subequations}
%%%%%%%%%%%%%%%%%%%%%%%%%%%%%%%%%%%%%%%%%%%%%%%%%%%%%%%%%%%%%%%%%%%%%%%%%%
with $\cot2\theta=(\Delta/2)/\Omega$ and $\Delta=\omega_{0}-\omega_{L}$. In 
this picture the force is given by the relation:
\begin{eqnarray}
\vec F = -i\vec k_{L}\hbar\Omega (R^{+}-R^{-}). \label{fo}
\end{eqnarray}
Here $R^{+}=|\bar 2\rangle \langle \bar 1|$ and $R^{-}=|\bar 1\rangle \langle \bar 2|$
are new quasispin operators operating in the dressed state picture. They obey the 
same commutation relations as the old ones. To obtain the explicit expression for the 
force we need the equations of motion for the new dressed-state operators. Therefore 
we write the Hamiltonian in the dressed-state representation:
\begin{eqnarray}
H&=&\sum_{k}\hbar(\omega_{k}-\omega_{L})a^{\dagger}_{k}a_{k} + \hbar \bar \Omega R_{z} \nonumber \\
&+& i\sum_{k}\{(\sin2\theta R_{z}/2 + \cos^{2}\theta R^{-}-\sin^{2}\theta R^{+})a^{\dagger}_{k} \nonumber \\
&-& H.c.\}. \label{DHm}
\end{eqnarray}
Here $R_{z}=|\bar 2\rangle \langle \bar 2|-|\bar 1\rangle \langle \bar 1|$ is the dressed-state inversion
operator while $\bar \Omega = \sqrt{\Omega^{2}+(\Delta/2)^{2}}$ is the generalized Rabi frequency.
The Heisenberg equation for an arbitrary dressed-state atomic operator $Q$ is:
\begin{eqnarray}
\frac{d}{dt}\langle Q(t)\rangle = \frac{i}{\hbar}\langle[H,Q(t)]\rangle. \label{EqH}
\end{eqnarray}
Here the notation $\langle \cdots \rangle$ indicates averaging over the initial state of both the atoms 
and the EMF environmental reservoir.

Introducing the Hamiltonian (\ref{DHm}) in Eq.~(\ref{EqH}) one arrives at:
\begin{eqnarray}
&{}&\frac{d}{dt}\langle Q\rangle - i\bar \Omega\langle [R_{z},Q]\rangle = \nonumber \\
&-&\sum_{k}\frac{(\vec g_{k}\cdot \vec d)}{\hbar}\langle a^{\dagger}_{k}[\sin2\theta R_{z}/2 + \cos^{2}\theta R^{-}-\sin^{2}\theta R^{+},Q]\rangle \nonumber \\
&+& H.c., \label{EqHt}
\end{eqnarray}
where for the in general non-Hermitian atomic operators $Q$, the H.c. terms should be
evaluated without conjugating $Q$, i.e., by replacing $Q^{+}$ with $Q$ in the Hermitian 
conjugate parts. Assuming that the atomic subsystem couples weakly to the surrounding EMF, 
i.e., in the bad-cavity limit, the EMF operators can be eliminated from the above equation of
motion, Eq.~(\ref{EqHt}). On solving formally the Heisenberg equations for the EMF field operators 
one can represent the solutions in the form:
\begin{eqnarray}
a^{\dagger}_{k}(t)&=&a^{\dagger}_{k}(0)e^{i\bar \Delta_{k}t} + \pi\frac{(\vec g_{k}\cdot \vec d)}{\hbar}
\{\sin{2\theta}R_{z}(t)\delta(\omega_{k}-\omega_{L})/2 \nonumber \\
&+& \cos^{2}\theta R^{+}(t)\delta(\omega_{k}-\omega_{L}-2\bar\Omega) \nonumber \\
&-& \sin^{2}\theta R^{-}(t)\delta(\omega_{k}-\omega_{L}+2\bar\Omega)\}, \label{Eqf}
\end{eqnarray}
where $\bar \Delta_{k} = \omega_{k}-\omega_{L}$ and $a_{k}=[a^{\dagger}_{k}]^{\dagger}$.
Here the contributions leading to a small Lamb shift were ignored.
Substituting Eq.~(\ref{Eqf}) in Eq.~(\ref{EqHt}) and summing over $k$ one arrives at 
the following master equation:
\begin{eqnarray}
&\frac{d}{dt}\langle Q\rangle - i\bar \Omega\langle[R_{z},Q]\rangle = \nonumber \\
&- \langle(\gamma_{0}\sin2\theta R_{z}/2 + \gamma_{+}\cos^{2}\theta R^{+} - \gamma_{-}\sin^{2}\theta R^{-})\nonumber \\
&\times[\sin2\theta R_{z}/2 + \cos^{2}\theta R^{-} - \sin^{2}\theta R^{+},Q]\rangle + H.c. \label{ME}
\end{eqnarray}
Here, in free space, $\gamma(\omega)=2d^{2}\omega^{3}/(3\hbar c^{3})$ and for $\gamma_{0}$ we have 
$\omega \equiv \omega_{L}$ while for $\gamma_{\pm}$ we have $\omega = \omega_{L} \pm 2\bar \Omega$.
Note that for usual vacuum modes and moderate driving, i.e. $\bar \Omega/\omega_{L} \to 0$, we have $\gamma_{0}=\gamma_{+}=\gamma_{-}$. We anticipate that, for modified environmental reservoirs such as 
low quality optical cavities, or for very intense driving in free space with $\bar \Omega/\omega_{L} \ll 1$ 
but not zero, this is not the case, namely, $\gamma_{0}\not=\gamma_{+}\not=\gamma_{-}$ \cite{CHK}. 

We emphasize here that the master equation (\ref{ME}) describes also, under the Born-Markov 
conditions, the case of a driven two-level particle that is damped by a modified reservoir, 
i.e., when the density of electromagnetic field modes is different at various dressed-states 
transitions. In this case $\gamma_{\pm} \propto g(\omega_{L}\pm 2\bar \Omega)$ while 
$\gamma_{0} \propto g(\omega_{L})$, where $g(\omega)$ characterizes the atom-environment 
coupling strength \cite{melk,lsq,john,gxl,rev,CHK}. In the next subsections, we shall obtain the equations 
of motion as well as the expression for the optical force acting on a strongly driven two-level 
particle in various environments.

%%%%%%%%%%%%%%%%%%%%%%%%%%%%%%%%%%%%%%%%%%%%%%%%%%%%%%%%%%%%%%%%%%%%%%%%%%
\subsection{Equations of motion}
%%%%%%%%%%%%%%%%%%%%%%%%%%%%%%%%%%%%%%%%%%%%%%%%%%%%%%%%%%%%%%%%%%%%%%%%%%
Using Eq.~(\ref{ME}) one can obtain the equations of motion for the 
operators of interest:
%%%%%%%%%%%%%%%%%%%%%%%%%%%%%%%%%%%%%%%%%%%%%%%%%%%%%%%%%%%%%%%%%%%%%%%%%%
\begin{subequations}
\label{EqM}
\begin{align}
&\frac{d}{dt}\langle R_{z}\rangle = -2\gamma_{p}\langle R_{z}\rangle + \gamma_{0}\sin{4\theta}
\langle R^{+} + R^{-}\rangle/2 - 2\gamma_{m}, \\
&\frac{d}{dt}\langle R^{+}\rangle = \langle R^{+}\rangle[2i\bar \Omega - (\gamma_{0}\sin^{2}2\theta + \gamma_{p})] 
\nonumber \\
&-(\gamma_{+}+\gamma_{-})\sin^{2}2\theta\langle R^{-}\rangle/4 + \gamma_{f} \nonumber \\
&+\sin{2\theta}\langle R_{z}\rangle(\gamma_{+}\cos^{2}\theta - \gamma_{-}\sin^{2}\theta)/2, \\
&\frac{d}{dt}\langle R^{-}\rangle = -\langle R^{-}\rangle[2i\bar \Omega + (\gamma_{0}\sin^{2}2\theta + \gamma_{p})] 
\nonumber \\
&-(\gamma_{+}+\gamma_{-})\sin^{2}2\theta\langle R^{+}\rangle/4 + \gamma_{f} \nonumber \\
&+\sin{2\theta}\langle R_{z}\rangle(\gamma_{+}\cos^{2}\theta - \gamma_{-}\sin^{2}\theta)/2. 
\end{align}
\end{subequations}
%%%%%%%%%%%%%%%%%%%%%%%%%%%%%%%%%%%%%%%%%%%%%%%%%%%%%%%%%%%%%%%%%%%%%%%%%%
Here $\gamma_{p}=\gamma_{+}\cos^{4}\theta + \gamma_{-}\sin^{4}\theta$, 
$\gamma_{m}=\gamma_{+}\cos^{4}\theta - \gamma_{-}\sin^{4}\theta$ while 
$\gamma_{f}=\sin2\theta(\gamma_{0}+\gamma_{+}\cos^{2}\theta+\gamma_{-}\sin^{2}\theta)/2$.
Further, $\cos^{2}\theta=(1+(\Delta/2)/\bar \Omega)/2$, $\sin^{2}\theta=(1-(\Delta/2)/\bar \Omega)/2$
and $\sin{2\theta}=\Omega/\bar \Omega$.

%%%%%%%%%%%%%%%%%%%%%%%%%%%%%%%%%%%%%%%%%%%%%%%%%%%%%%%%%%%%%%%%%%%%%%%%%%
\subsection{Optical force}
%%%%%%%%%%%%%%%%%%%%%%%%%%%%%%%%%%%%%%%%%%%%%%%%%%%%%%%%%%%%%%%%%%%%%%%%%%
Substituting the steady-state solution of Eq.~(\ref{EqM}) in Eq.~(\ref{fo}) we 
obtain the following mean expression for the optical force:
\begin{eqnarray}
\langle F\rangle = k_{L}\hbar\frac{2\bar \gamma\Omega\bar \Omega\sin{2\theta}}{4\gamma_{p}\bar\Omega^{2} +
\gamma_{1}(\gamma_{2}\gamma_{p}-\gamma_{0}\tilde \gamma\sin2\theta\sin4\theta/4)}. \label{F}
\end{eqnarray}
where $\bar \gamma = \gamma_{+}\cos^{4}\theta(\gamma_{0}+2\gamma_{-}\sin^{2}\theta)
+\gamma_{-}\sin^{4}\theta(\gamma_{0}+2\gamma_{+}\cos^{2}\theta)$, 
$\gamma_{1}=\gamma_{0}\sin^{2}2\theta + \tilde \gamma\cos2\theta$, 
$\gamma_{2}=\gamma_{0}\sin^{2}2\theta + \gamma_{+}\cos^{2}\theta+\gamma_{-}\sin^{2}\theta$ and 
$\tilde \gamma = \gamma_{+}\cos^{2}\theta - \gamma_{-}\sin^{2}\theta$.

In the following section we shall analyze the optical force acting on a two-level 
particle in a running wave laser field in more details.

%%%%%%%%%%%%%%%%%%%%%%%%%%%%%%%%%%%%%%%%%%%%%%%%%%%%%%%%%%%%%%%%%%%%%%%%%%%
\section{Results and discussions}
%%%%%%%%%%%%%%%%%%%%%%%%%%%%%%%%%%%%%%%%%%%%%%%%%%%%%%%%%%%%%%%%%%%%%%%%%%%
In the case when the driven atom is surrounded by the usual vacuum modes 
and $2\bar \Omega/\omega_{L} \to 0$, the spontaneous decay rates corresponding 
to different dressed-state transitions are equal, i.e. 
$\gamma_{0}=\gamma_{+}=\gamma_{-}\equiv \gamma$. This will lead to the well-known 
expression for the optical force, that is:
\begin{eqnarray}
\langle F\rangle = 2k_{L}\hbar \gamma \frac{\Omega^{2}}{\Delta^{2}+\gamma^{2}+2\Omega^{2}}.
\label{FV}
\end{eqnarray}
For strong fields, i.e. $\Omega^{2} \gg \Delta^{2}+ \gamma^{2}$,  one arrives at the maximal 
expression of the force:
\begin{eqnarray}
\langle F\rangle = k_{L}\hbar \gamma. \label{FMax} 
\end{eqnarray}
Notice, that the particle velocity $v$ can be included in Eq.~(\ref{FV}) via the modified 
detuning, i.e., $\Delta \to \Delta - k_{L}v$.
%%%%%%%%%%%%%%%%%%%%%%%%%%%%%%%%%%%%%%%%%%%%%%%%%%%%%%%%%%%%%%%%%%%%%%%%%%
\begin{figure}[t]
\includegraphics[width=8cm]{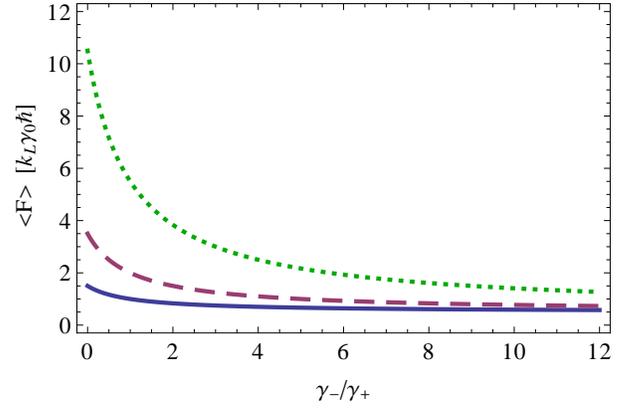}
\caption{\label{fig-2} (color online) The maximal optical force acting on a two-level 
particle in a modified environment. The solid line is for $\gamma_{-}/\gamma_{0}=1$,
the dashed line corresponds to $\gamma_{-}/\gamma_{0}=3$ while the dotted ones to
$\gamma_{-}/\gamma_{0}=10$. }
\end{figure}
%%%%%%%%%%%%%%%%%%%%%%%%%%%%%%%%%%%%%%%%%%%%%%%%%%%%%%%%%%%%%%%%%%%%%%%%%%%

In the following we shall obtain the corresponding expression for the maximal force 
in the strong-field limit and in the modified reservoirs. In the intense-field 
limit, that is $\Omega \gg \{\Delta, \gamma_{\pm},\gamma_{0}\}$, we have 
$\cos^{2}\theta \approx \sin^{2}\theta \to 1/2$ while $\sin2\theta \to 1$.
Therefore, the optical force acting on a strongly driven two-level atom in a modified 
reservoir is:
\begin{eqnarray}
\langle F\rangle=k_{L}\hbar\gamma_{0}\biggl(1/2 + 
\frac{\gamma_{+}\gamma_{-}}{\gamma_{0}(\gamma_{+}+\gamma_{-})}\biggr ).
\label{FME}
\end{eqnarray}
One can observe that the optical force in the strong-field limit depends on the 
spontaneous decay rates at particular dressed-state transitions. By modifying 
of these dressed-decay rates via a suitable environment reservoir one can 
influence the magnitude of the optical force. In particular, when 
$\gamma_{+} \gg \gamma_{-}$  we have for the optical force the following 
expression 
\begin{eqnarray}
\langle F\rangle=k_{L}\hbar\gamma_{0}\biggl(1/2 + 
\frac{\gamma_{-}}{\gamma_{0}}\biggr ).
\label{FMEM}
\end{eqnarray}
Now if $\gamma_{-} \gg \gamma_{0}$ we obtain that the optical force is greater 
than $k_{L}\hbar\gamma_{0}$. Otherwise if $\gamma_{-} \ll \gamma_{0}$, 
we have $\langle F\rangle=k_{L}\hbar\gamma_{0}/2$.

Conversely, if $\gamma_{-} \gg \gamma_{+}$ we get for the 
force in Eq.~(\ref{FME}):
\begin{eqnarray}
\langle F\rangle=k_{L}\hbar\gamma_{0}\biggl(1/2 + 
\frac{\gamma_{+}}{\gamma_{0}}\biggr ),
\label{FMEMM}
\end{eqnarray}
which is again much larger than $k_{L}\hbar\gamma_{0}$ when $\gamma_{+} \gg \gamma_{0}$
or $\langle F\rangle=k_{L}\hbar\gamma_{0}/2$ if $\gamma_{+} \ll \gamma_{0}$. 
Finally, if $\gamma_{+}=\gamma_{-} \gg \gamma_{0}$ we again have an increase in the 
optical force. These features are shown in Figure (\ref{fig-2}) where Eq.~(\ref{FME}) 
is plotted.

Further, we shall evaluate the maximal optical force when the two-level emitter 
is pumped with a very intense laser field in free space such that 
$2\bar \Omega/\omega_{L} \ll 1$ but not zero. In this case the dressed decay rates can 
be determined as follows: 
\begin{eqnarray}
\gamma_{\pm}=\gamma_{0}(1 \pm 2\bar \Omega/\omega_{L})^{3}. 
\label{ddr}
\end{eqnarray}
Substituting these expressions in Eq.~(\ref{FME}) and to the second order in the small 
parameter $\bar \Omega/\omega_{L}$ one arrives at:
\begin{eqnarray}
\langle F\rangle=k_{L}\hbar\gamma_{0}\biggl(1 - 3(2\bar \Omega/\omega_{L})^{2}\biggr ).
\label{FFS}
\end{eqnarray}
One can observe here that the maximal force in free space is smaller for very intense 
driving than for moderate pumping (see Eq.~\ref{FMax}). In particular, if $2\bar \Omega/\omega_{L}=0.1$
then the force in Eq.~(\ref{FFS}) is three procents smaller than the force (\ref{FMax}). 
This result is somehow counterintuitive as one may expect the force to be larger for higher 
field intensities.

%%%%%%%%%%%%%%%%%%%%%%%%%%%%%%%%%%%%%%%%%%%%%%%%%%%%%%%%%%%%%%%%%%%%%%%%%%%
\section{Summary}
%%%%%%%%%%%%%%%%%%%%%%%%%%%%%%%%%%%%%%%%%%%%%%%%%%%%%%%%%%%%%%%%%%%%%%%%%%%
In summary, we have investigated the optical force acting on a two-level atom 
in a running wave laser and in a modified surrounding electromagnetic field 
reservoir. For this, we obtained the master equation describing this process
and, correspondingly, we obtained the equations of motion for the dressed-state 
operators of interest which helped to get the optical force. The obtained optical 
force shows a strong dependence on the modifed electromagnetic reservoir via the 
dressed decay rates at particular frequencies. In particular, it can be much larger 
than the corresponding force in the free space. Finally, we evaluated the maximal 
optical force acting on a two-level emitter in very intense laser fields and in the 
free space.

\begin{acknowledgments}
We acknowledge valuable discussions with Christoph H. Keitel.
\end{acknowledgments}

%%%%%%%%%%%%%%%%%%%%%%%%%%%%%%%%%%%%%%%%%%%%%%%%%%%%%%%%%%%%%%%%%%%%%%%%%%%%%%%%%%%%%%%
{\small $^\dagger$ On leave from \it{Institute of Applied Physics, Academy of Sciences of Moldova, 
Academiei str. 5, MD-2028 Chi\c{s}in\u{a}u, Moldova.}}
%%%%%%%%%%%%%%%%%%%%%%%%%%%%%%%%%%%%%%%%%%%%%%%%%%%%%%%%%%%%%%%%%%%%%%%%%%%%%%%%%%%%%%%
%%%%%%%%%%%%%%%%%%%%%%%%%%%%%%%%%%%%%%%%%%%%%%%%%%%%%%%%%%%%%%%%%%%

%%%%%%%%%%%%%%%%%%%%%%%%%%%%%%%%%%%%%%%%%%%%%%%%%%%%%%%%%%%%%%%%%%%%%%%%%%%%%%%%%%%%%%%%%%%
\end{document}